\documentstyle[12pt]{article}

\begin{document}
{\pagestyle{empty}
\rightline{May 1994}
\vspace{10mm}

\centerline{\Large \bf Gravitational Force by Point Particle }
\centerline{\Large \bf in Static Einstein Universe }
\vspace{10mm}

\centerline{Masakatsu Kenmoku \footnote{E-mail address:
kenmoku@jpnyitp.bitnet}, Kaori Otsuki and Rieko Sakai}
\centerline{{\it Department of Physics} }
\centerline{{\it Nara Women's University, Nara 630, Japan} }

\vspace{10mm}

\vspace{10mm}

\centerline{\bf Abstract}
\vspace{10mm}

The gravitaional force produced by a point particle, like the sun,
 in the background of the static Einstein universe is studied.
Both the approximate solution in the weak field limit
and exact solution are obtained.
The main properties of the solution are
{\it i}) near the point particle,
the metric approaches the Schwarzschild one and
the radius of its singularity becomes larger
than that of the Schwarzschild singularity,
{\it ii}) far from the point particle,
the metric approaches the static Einstein closed universe.
The maximum length of the equator of the universe becomes smaller
than that of the static Einstein universe
due to the existence of the point particle.
These properties show the strong correlation betweem the particle
and the universe.

\vspace{10mm}
\vspace{10mm}

\hfil
\vfill
\newpage}

\vspace{1.5cm}

\noindent
{\large \bf 1.\ \ Introduction}


\vspace{0.5cm}

Since originally Einstein introduced  cosmological constant term in his
gravitational field equations to obtain the static universe,
the existence of the cosmological constant has been paid noticeable
attention.
The observed cosmological constant is many order of magnitude smaller
than the theoretically expected value.
This is known as the cosmological constant problem
 \cite{Wein89}, \cite{Caroll}.
Recently, it was reported that models of cosmology
with nonzero cosmological constant
can successfully account for the deep galaxy counts \cite{Fukugita}
and the observed flat rotation curve of galaxies \cite{Suto}.
Independently of these, recent observation of the Hubble parameter
strongly suggests the existence of the positive value of
the cosmological constant \cite{Fuku2}.
 It has been widely investigated to apply the Einstein equations
with a cosmological term to obtain a gravitational field produced
by a mass
density and to study a whole universe \cite{Wein3},
\cite{Ueha}.

In this paper, we apply the Einstein equations
with a cosmological term to the general isotropic metric
and derive the gravitational potential produced by the spherically
symmetric point matter in the closed Einstein universe.
{}From another point of view of this problem, we study the effect of
the existence of the background universe to the gravitational potential
produced by a point matter.
Nature of the closed universe may affect strongly
to the local gravitational potential.
The correlation between the local quantity
( the gravitational potential by the point matter)
and the global quantity
(the static Einstein universe) are examined.

The paper is organized as follows. In Sec.2 the Einstein
 equations with the cosmological term and the constraint
on energy density and the cosmological constant
of our model are presented. In Sec.3 we study
the approximate solutions of weak static field
in order to understand the physical meaning well.
In Sec.4 we consider
the exact solution. Concluding discussions are given in Sec.5.


\vspace{1.5cm}

\noindent
{\large \bf 2. Einstein Equations and  Constraint on Energy Density}

\vspace{0.5cm}

We start with the Einstein equations with comological term
$\Lambda$,
\begin{equation}
R_{\mu \nu}-{1 \over 2} R g_{\mu \nu} - \Lambda g_{\mu \nu}
= -8 \pi G~T_{\mu \nu}~, \label{eq1}
\end{equation}
where $ G $ is the gravitational constant and $ T_{\mu \nu} $
is the energy-momentum tensor of the matter.
Throughout this paper we follow the conventions of Ref. \cite{Wein72}.
We now apply the theory to a static isotropic metric
in order to represent the gravitational fields produced by
a point particle in the background of the static universe.
The general form of the static isotropic metric in the standard form is
\begin{equation}
ds^2 = - e^{\nu} dt^2 + e^{\lambda} dr^2 + r^2
\{d \theta^2 + \sin^2 {\theta} d \phi^2 \}~, \label{eq2}
\end{equation}
where $\nu$ and $\lambda$ are functions of radial coodinate $r$ only.

The energy-momentum tensor is assumed to
be composed of two parts, i.e.,
that of the parfect fluid $ T_F^{\mu \nu} $ and
that of a point particle
$ T_P^{\mu \nu}$:
\begin{eqnarray}
T^{\mu \nu} &=& T_F^{\mu \nu} + T_P^{\mu \nu}~, \label{eq301} \\
T_F^{\mu \nu} &=& p(r) g^{\mu \nu}
+ ({\rho}(r) + p(r)) u^{\mu}u^{\nu}~,
\label{eq302} \\
T_P^{\mu \nu} &=& M {\delta}^{(3)}(\vec{r})~, \label{eq303}
\end{eqnarray}
where $\rho$ , $p$ and $u^{\mu}$
denote the energy density , the pressure and the four-velocity
of the perfect fluid respectively
and $M$ denotes the mass of the particle.
The energy-momentum tensor obeys the local conservation equation,
\begin{equation}
 \nabla _\mu T^{\mu \nu} = 0~.   \label{eq4}
\end{equation}

Using the metric Eq. (\ref{eq2}) and the energy-momentum tensor
Eqs. (\ref{eq301})- (\ref{eq303}),
the independent component of the Einstein equations are
\begin{eqnarray}
{1 \over r^2} - e^{- \lambda }
({1 \over r^2} - {\lambda ' \over r} ) -
\Lambda &=& 8 \pi G \{ \rho
+ M \delta^{3}(\vec{r}) \}~, \label{eq501} \\
{1 \over r^2} - e^{- \lambda} ( {1 \over r^2} + {\nu ' \over r} ) -
\Lambda &=& - 8 \pi G p~, \label{eq502} \\
{1 \over 2} e^{- \lambda} \{ {1\over 2} \nu ' \lambda ' - \nu '' -
{1 \over 2} \nu '^2 - {\nu ' \over r} +{ \lambda ' \over r} \} -
\Lambda &=& - 8 \pi G p~,  \label{eq503}
\end{eqnarray}
where prime stands for derivative with respect to $r$.
The local conservation of the energy-momentum tensor becomes
\begin{equation}
p' + {\nu ' \over 2} \{ p + \rho + M \delta ^{(3)} (\vec{r}) \} = 0~.
\label{eq6}
\end{equation}

Here we impose the constraint between the energy density and
the cosmological constant, in order to obtain the solution of
the gravitational potential in the static Einstein universe, as
\begin{equation}
\rho = \rho _0 \quad ({\rm constant})~, \label{eq701}
\end{equation}
and
\begin{equation}
\Lambda = 4 \pi G \rho _0 = {1 \over a^2}~, \label{eq702}
\end{equation}
where $a$ is a kind of the constant radius of the universe.

With the constraint Eqs.(\ref{eq701}) and (\ref{eq702}),
and using the dimensionless radial variable
$$ \eta = r/a~, $$
the Einstein equations Eqs.(\ref{eq501})-(\ref{eq503})
 and the local conservation of the energy-momentum tensor
Eq.(\ref{eq6}) become as
\begin{eqnarray}
{1 \over \eta^2} - e^{-\lambda} ( {1 \over \eta^2} -
{ \dot{\lambda} \over \eta} ) - 3 &=&
8 \pi a^2 G M \delta ^{(3)} (\vec{r})~,
\label{eq901} \\
{1 \over \eta^2} - e^{-\lambda} ( {1 \over \eta^2} +
{\dot{\nu} \over \eta} ) -1 &=& - 8 \pi a^2 G p~, \label{eq902} \\
{1 \over 2} e^{- \lambda} \{ {1 \over 2} \dot{\nu} \dot{\lambda}
 - \ddot{\nu} - {1 \over2} \dot{\nu}^2
-{ \dot{\nu} \over \eta} + {\dot{\lambda} \over \eta} \} - 1
&=& - 8 \pi a^2 G p~, \label{eq903}
\end{eqnarray}
and
\begin{equation}
\dot{p} + {\dot{\nu} \over 2} \{ p + \rho _0
+ M \delta ^{(3)} (\vec{r}) \} =0~, \label{eq10}
\end{equation}
where dot stands for derivative with respect to $\eta$. Among these
four equations Eqs.(\ref{eq901})-(\ref{eq903}) and (\ref{eq10}),
three independent equations determine
the three unknown functions ${\nu}$, $\lambda$ and $p$ .

 The boundary condition is required, in order to represent
 the Newtonian potential in the Einstein universe, as \\
{\it i}) near the point particle,
the solution represents the Newtonian potential, \\
{\it ii}) far from the point particle,
the metric reduces to one of the Einstein universe.

 The exact solution of the vacuum Einstein
equation with cosmological term
 is known as the Schwarzschild-de Sitter solution:
$$ e^{\nu} = e^{-\lambda}
= 1 - {2 G M \over r} + {\Lambda \over 3} r^2~.
$$ \\
In the following sections, we will study another type of solution
of the Einstein equations with cosmological term
not in the flat space-time but in the closed universe.


\vspace{1.5cm}

\noindent
{\large \bf 3. Weak Field Approximation}

\vspace{0.5cm}
In this section, we study an approximate solution of
Eqs.(\ref{eq901})-(\ref{eq903}) and (\ref{eq10}),
which is of first order in the  gravitational constant $G$.
This weak field solution plays the important role to understand
the physical meaning of the gravitational potential
produced by the point particle
in the background of the closed universe.
The metric tensor is expressed as
\begin{eqnarray}
-g_{00} = e^{\nu} = 1 + 2 \phi (\eta)~, \label{eq1101} \\
 g_{11} = e^{\lambda} = {{ 1 + 2 \psi (\eta) } \over {1 -\eta ^2}}~.
 \label{eq1102}
\end{eqnarray}
The weak fields $\phi$ and $\psi$ as well as the pressure $p$
are meaningful only within the linear approximation.
{}From the local conservation law Eq.(\ref{eq10}), the pressure is
expressed by $\phi$ as,
\begin{equation}
p = - \rho _0 \phi~. \label{eq1103}
\end{equation}
Eliminating the pressure using Eq.(\ref{eq1103}),
the Einstein fiels equations in Eqs.
(\ref{eq901})- (\ref{eq903}) become,
\begin{eqnarray}
{(1 -\eta ^2 ) \over \eta} \dot{\psi} + ( {1 \over \eta ^2} - 3 ) \psi
 &=& 4 \pi a^2 G M \delta^{(3)} (\vec{r})~, \label{eq1201} \\
\eta \dot{\phi} + {\eta ^2 \over{ 1 - \eta^2 }}
\phi &=& \psi~, \label{eq1202}
\end{eqnarray}
and
\begin{equation}
\eta \ddot{\phi} + {1 \over{1 - \eta ^2}} \dot{\phi}
 + {2 \eta \over (1 - \eta ^2)^2} \phi = \dot{\psi}~. \label{eq1203}
\end{equation}
  Eq. (\ref{eq1203}) is atomatically satisfied by using
Eqs. (\ref{eq1201}) and (\ref{eq1202}).
We can solve Eq.(\ref{eq1201}) for $ \psi $ as
\begin{equation}
\psi = {\psi_0 \over {\eta (1 - \eta^2)}}~, \label{eq13}
\end{equation}
where $ \psi_0 $ is a integration constant.
Inserting Eq.(\ref{eq1202}) into Eq.(\ref{eq1201}),
the eqation for $\phi$ is obtained
\begin{equation}
\triangle \phi + {3 \over a^2} \phi
= 4 \pi G M \delta^{(3)} (\vec{r})~,
\label{eq14}
\end{equation}
where the Laplacian operator in static closed universe is
\begin{eqnarray}
\triangle &\equiv& {1 \over \sqrt{g}} \partial _i
( g^{ij} \sqrt{g} \partial _j )~,  \nonumber \\
&=& {1 \over a^2} [ ( 1 - \eta ^2 )
{\partial^2 \over{\partial \eta^2}} +
{(2 - 3 \eta^2)  \over \eta} { \partial \over{\partial \eta}}]~.
\label{eq15}
\end{eqnarray}
The general solution of this equation for $\phi$ is
\begin{eqnarray}
\phi = b_1 {1-2\eta^2 \over{\eta}} + b_2 \sqrt{1 - \eta^2}~,
\label{eq16}
\end{eqnarray}
where $b_1$ and $b_2$ are integration constants.
According to boundary  condition, which is in the previous section,
 the integration constants are determined as
$$ \psi_0 = - b_1 = G M / a ~, \quad  b_2 = 0 ~. $$

The invariant distance takes
\begin{eqnarray}
ds^2 = - \{ 1 - {2 G M \over r} (1-(r/a)^2)\} dt^2
+  { 1 + {\displaystyle {2GM \over{r(1-(r/a)^2)}}}
 \over{1-(r/a)^2}} dr^2
+ r^2 \{d\theta^2 + \sin^2 \theta d\phi ^2\}~.  \label{eq17}
\end{eqnarray}
The comoving radial coodinate, in the linear approximation, is given by
\begin{equation}
\chi (r) = \int_{0}^{r}
{dr \over{\sqrt{1- (r/a)^2}}} = a~\sin^{-1} (r/a).
\label{eq18}
\end{equation}
The gravitational potential is expressed
by using the comoving radial coodinate
\begin{eqnarray}
\phi ( \rho) &=& -{{ GM \cos(2\chi/a)} \over {a~ \sin(\chi/a)}}~,
\nonumber \\
            &\simeq& -GM/\chi \quad for \quad \chi/a \ll 1 ~,
\label{eq19}
\end{eqnarray}
and has the symmetry
\begin{equation}
g_{00}(\pi - \chi /a) = g_{00} (\chi /a)~. \label{eq20}
\end{equation}
This means that the gravitational potential,
as well as the mass of the point particle,
is symmetric with respect to the north pole ($ \chi \simeq 0 $)
and the south pole ($ \chi \simeq a \pi $) of the closed universe.

So far, we studied some physical implications of our model within
the weak field approximation. In the next section,
we will explore the exact solution, which will clear the non-linear
effects of our model.


\vspace{1.5cm}

\noindent

{\large \bf 4. Exact Solution}

\vspace{0.5cm}

In this section, we discuss exact solution of
Eqs.(\ref{eq901})-(\ref{eq903}) and (\ref{eq10}), which are
the Einstein equations with cosmological term Eq.(\ref{eq1})
in the general static isotropic metric Eq.(\ref{eq2})
with the constraint Eqs.(\ref{eq701}) and (\ref{eq702}).
The exact solution is obtained in a similar way as the approximate one.
The local energy-momentum conservation Eq.(\ref{eq10}) is solved and
the pressure is obtained as
\begin{equation}
p = \rho_0 (e^{-\nu/2} -1)~, \label{eq21}
\end{equation}
where integration constant is choosen to coincide with
the approximate solution Eq.(\ref{eq1103}) in the weak field limit.
The radial component of the metric is solved
from the Eq. (\ref{eq901}) as
\begin{eqnarray}
e^{\lambda} = {1 \over {1 - \eta ^2 - b / \eta}}~, \label{eq22}
\end{eqnarray}
 where  $b$ is the integration constant.
Inserting Eqs. (\ref{eq21}) and (\ref{eq22}) into Eq. (\ref{eq902}),
the equation for the time component of the metric becomes
\begin{equation}
{d \over{d \eta}} e^{\nu/2} +
{{\eta - b/(2\eta^2)} \over {1 - \eta ^2 - b/\eta}} e^{\nu/2}
 = {\eta \over { 1 - \eta^2 - b/\eta}}. \label{eq23}
\end{equation}
The solution is obtained in the integral form:
\begin{equation}
e^{\nu/2} = 1 +
{b \over 2} \sqrt{1 - \eta^2 - b/\eta}
\int_{\eta_{\ast}}^{\eta} {1 \over {\eta^2 ( 1-\eta^2 - b/\eta)^{3/2}}}
 d \eta, \label{eq24}
\end{equation}
where the lower value of the integration
$\eta_{\ast}$ is recognized as a integration constant.
The integration in Eq.(\ref{eq24}) is  a kind of the elliptic integral
and is estimated approximately in extreme cases. The resultant
asymptotic expression for the time component of the metric is
\begin{eqnarray}
e^{\nu /2} \simeq \left\{
\begin{array}{l}
1 \ \ \ for \ \ \  b = 0 ~,\\
1 - b {\displaystyle {(1- 2\eta^2) \over {2\eta}}} +
b {\displaystyle {{( 1-2 \eta^2_{\ast})} \over {2 \eta_{\ast}}}
 {\sqrt{1-\eta^2} \over  { \sqrt{1-\eta^2_{\ast}}}}}
\ \ \ for \ \ \  b\ll 1 ~,\\
{\displaystyle {\sqrt{1-b/\eta} \over \sqrt{1-b/\eta_{\ast}}}}
\ \ \ for \ \ \  a \rightarrow \infty ~.
\label{eq2401}
\end{array}
\right.
\end{eqnarray}

We impose the same boundary condition as in the previous section,
and one integration constant is determined as
 \begin{equation}
b = 2 G M / a~. \label{eq2501}
\end{equation}
 Another integration constant $\eta_{\ast}$ is
determined as the solution
of the condition
\begin{equation}
{\partial \over {\partial \chi}} g_{00}
 = \sqrt{1-\eta^2-b/\eta} {\partial \over {\partial \eta}} g_{00}
 = 0~, \nonumber
\end{equation}
at the equator of the closed universe ( which corresponds to the
largest value of the radial coodinate $ \eta_+ $ defined in
Eq.(\ref{eq2601}) below)
and, where, $\chi$ denotes the comoving radial coodinate.
We note that the above condition maches the boundary condition.
This integration constant is estimated, in a weak field approximation
 ($ b \ll 1 $), as
\begin{equation}
\eta_{\ast} \simeq 1 / \sqrt{2}~. \label{eq2502}
\end{equation}

{}From the asymptotic expression of the time component of the metric,
 three specific features are obtained;
{\it i}) if the mass of the point particle
disappears ($b = 2GM/a =0$),
the solution reduces to  one of the Einstein universe
as in the case of the approximate solution,
{\it ii}) for small $b$~ ($ = 2 G M / a  \ll 1$),
it reproduces the approximate solution of Eq.(\ref{eq16})
in the previous section ,
{\it iii}) if  the radius of the universe tends to infinity
( $ a \rightarrow \infty $),
it coincides with the Schwarzschild solution.

Three pole positions of the radial part of the metric, defined as
\begin{eqnarray}
(g_{11})^{-1} &\equiv& 1 - \eta^2 - b/\eta ~, \nonumber \\
              &=&  - {1 \over \eta}
(\eta - \eta_{-})(\eta - \eta_0)(\eta - \eta_{+})~, \quad
( \eta_{-} < \eta_0 < \eta_{+}) , \label{eq2601}
\end{eqnarray}
are obtained for small $b = 2GM/a$ as
\begin{equation}
\eta_{-} \simeq - 1 - b/2~, \quad
 \eta_0 \simeq b + b^3~, \quad
 \eta_{+} \simeq 1 - b/2~.
 \label{eq2602}
\end{equation}
The following properties are obtained
from consideration of the pole positions.
(1) The radius of the singurarity by the mass of the particle
$ \eta_{0} $ becomes a little larger than
that of the Schwarzschild singularity $ b  =  2GM/a $.
(2) The maximum proper length $ a\eta_{+}$ becomes smaller than
that of the Einstein static universe.
As a result ,
the size of the universe around the equator ($ r = a\eta_{+},
\theta = \pi /2 $) is calculated to be ;
\begin{equation}
\int_{0}^{2\pi} \sqrt{g_{33}} d\phi =
\int_{0}^{2\pi} r \sin \theta d\phi =
2 \pi a \eta_{+}, \nonumber
\end{equation}
which is smaller than that of the Einstein universe.
The correlation between the part (the point particle)
and the whole (the universe) is a specific property of our solution.


\vspace{1.5cm}

\noindent

{\large \bf 5. Discussions}

\vspace{0.5cm}

We have obtained the approximate and the exact solution of
the spherically symmtric gravitational field produced
by the point particle
in the background of the static Einstein universe.
About this solution, the following points are noticed. \\
(a) The three dimensional scalar curvature is constant,
but the radial part of the metric has singularity
at the origin of the radial coodinate,
which corresponds to the energy source of the point
particle.\\
(b) The boundary condition of our solution at infinity
is not flat space-time. So  our solution does not agree with
the Schwarzschild solution contrary to the Birkhoff theorem
\cite{Birk}.
The Schwarzschild solution is realized only in the asymptotic
 case ( $ a \rightarrow  \infty $ ) in our solution
( see Eq. (\ref{eq2401})). \\
(c) As is well known, the Einstein universe solution is unstable.
This property reflects the presence of the  pressure
$ p = - \rho _0 \phi $ (Eq. (\ref{eq1103}))  or
$ p =  \rho_0 (e^{-\nu /2} - 1) $ (Eq. (\ref{eq21})) in our solution.
We have treated the energy density of the perfect fluid to be constant
$\rho=\rho_0$. If this energy density
is treated as a dynamical variable,
it will gather toward the position of the point particle
by the pressure and the universe will become unstable.

The work to obtain the gravitational
field produced by the various types
of the local quantities, i.e., mass, charge, etc.,
in the background of the expanding universe is
now under investigation.
Also, the topological structure of the singurarities of the metric
may be important \cite{Hawking} and that of our solution
will be discussed elsewhere.

\vspace{1.5cm}
\noindent
{\large \bf Acknowledgements}:
  We would like to thank K. Shigemoto for useful discussions and
a careful reading of this manuscript. We are also grateful to
K. Uehara for  useful discussions and
the estimation of the elliptic integral.

\vspace{1.0cm}
\newpage


\end{document}